\documentclass{article} 
\usepackage{nips12submit_e,times}
\usepackage{natbib}

\usepackage{amsmath}
\usepackage{amssymb}
\usepackage{tabularx}
\usepackage{url}

\usepackage{graphicx}
\usepackage{subfig}

\usepackage{pifont}
\newcommand{\yes}{\ding{52}\ }
\newcommand{\no}{\ding{56}\ }

\title{Theano: new features and speed improvements}

\author{
Fr\'ed\'eric Bastien, \rm \small \texttt{nouiz@nouiz.org}\\
\textbf{Pascal Lamblin,} \small \texttt{lamblinp@iro.umontreal.ca} \\
\textbf{Razvan Pascanu,} \small \texttt{r.pascanu@gmail.com} \\
\textbf{James Bergstra,} \small \texttt{james.bergstra@gmail.com} \\
\textbf{Ian Goodfellow,} \small \texttt{goodfeli@iro.umontreal.ca} \\
\textbf{Arnaud Bergeron,} \small \texttt{bergearn@iro.umontreal.ca} \\
\textbf{Nicolas Bouchard,} \small \texttt{nicolas.bouchard.1@gmail.com} \\
\textbf{David Warde-Farley,} \small \texttt{wardefar@iro.umontreal.ca} \\
\textbf{Yoshua Bengio,} \small \texttt{yoshua.bengio@umontreal.ca} \\
Dept. IRO, Universit\'e de Montr\'eal, Montr\'eal (QC), H3C 3J7, Canada
}

\nipsfinalcopy 

\begin{document}

\maketitle

\begin{abstract}

Theano is a linear algebra compiler that optimizes a user's
symbolically-specified mathematical computations to produce efficient low-level
implementations.  In this paper, we present new features and efficiency
improvements to Theano, and benchmarks demonstrating Theano's performance
relative to Torch7, a recently introduced machine learning library, and to
RNNLM, a C++ library targeted at recurrent neural networks.

\end{abstract}

\section{Introduction}

Theano was introduced to the machine learning community by
\citet{bergstra+al:2010-scipy} as a CPU and GPU mathematical compiler,
demonstrating how it can be used to symbolically define mathematical
functions, automatically derive gradient expressions, and compile these
expressions into executable functions that outperform implementations
using other existing tools.  \citet{bergstra+all-Theano-NIPS2011} then
demonstrated how Theano could be used to implement Deep Learning models.

In Section~\ref{sec:main_features}, we will briefly expose the main
goals and features of Theano. Section~\ref{sec:new_in_theano} will
present some of the new features available and measures taken to
speed up Theano's implementations.
Section~\ref{sec:benchmarks} compares Theano's performance with that
of Torch7~\citep{Torch-2011} on neural network benchmarks, and
RNNLM~\citep{Mikolov-Interspeech-2011} on recurrent
neural network benchmarks.
%

\section{Main features of Theano}
\label{sec:main_features}

Here we briefly summarize Theano's main features and advantages
for machine learning tasks.
\citet{bergstra+al:2010-scipy, bergstra+all-Theano-NIPS2011},
as well as Theano's
website\footnote{\url{http://deeplearning.net/software/theano/}} have
more in-depth descriptions and examples.

\subsection{Symbolic Mathematical Expressions}
\label{sec:symbolic_math}

Theano includes powerful tools for manipulating and optimizing graphs
representing symbolic mathematical expressions. In particular, Theano's 
\textit{optimization} constructs can eliminate duplicate or unnecessary
computations (e.g., replacing $x - x$ by $0$, obviating the need to compute
$x$ in the first place), increase numerical stability (e.g., by substituting
stable implementations of $\log(1+x)$ when
$x$ is tiny, or $\log(\operatorname{sigmoid}(x))$), or increase speed
(e.g., by using loop fusion to apply a sequence of scalar operations
to all elements of an array in a single pass over the data).

This graph representation also enables symbolic differentiation
of mathematical expressions, which allows users to quickly prototype
complex machine learning models fit by gradient descent without manually
deriving the gradient, decreasing the amount of code necessary and
eliminating several sources of practitioner error. Theano
now supports forward-mode differentiation via the R-operator
(see Section~\ref{sec:r-op}) as well as regular gradient
backpropagation. Theano is even able to derive symbolic
gradients through loops specified via the Scan operator (see
Section~\ref{sec:scan}).

\subsection{Fast to write and to execute}

Theano's dependency on NumPy and SciPy~\citep{scipy-2001} makes it easy to add an
implementation for a mathematical operation, leveraging the effort of
their developers, and it is always possible to add a more optimized
version that will then be transparently substituted where applicable.  For instance, Theano
defines operations on sparse matrices using SciPy's sparse matrix types
to hold values. Some of these operations simply call SciPy's functions,
other are reimplemented in C++, using BLAS routines for speed.

\subsection{Parallelism on the GPU}

Theano uses CUDA to define a class of $n$-dimensional (dense) arrays
located in GPU memory with Python bindings. Theano also includes CUDA
code generators for fast implementations of mathematical operations. Most
of these operations are currently limited to dense arrays of single-precision
floating-point numbers.

\subsection{Stability and community support}

Theano's development team has increased its commitment to code quality and
correctness as Theano usage begins to spread across university and industry
laboratories: a full test suite runs every night, with a shorter version
running for every pull request, and the project makes regular stable releases.
There is also a growing community of users who ask and answer questions every
day on the project's mailing lists.

\section{New features in Theano}
\label{sec:new_in_theano}

This section presents features of Theano that have been recently developed or
improved.  Some of these are entirely novel and extend the scenarios in which
Theano can be used (notably, Scan and the R operator); others aim at improving
performance, notably reducing the time not spent in actual computation (such as
Python interpreter overhead), and improving parallelism on CPU and GPU.

\subsection{Scan: Symbolic Loop in Theano}
\label{sec:scan}

Theano offers the ability to define symbolic loops through use of the
\emph{Scan Op},
a feature useful for working with recurrent models such as recurrent
neural networks, or for implementing more complex optimization algorithms such
as linear conjugate gradient.

Scan surmounts the practical difficulties surrounding other approaches
to loop-based computation with Theano. Using Theano's symbolically-defined
implementations within a Python loop prevents symbolic differentiation through
the iterative process, and prevents certain graph optimizations from being applied.
Completely unrolling the loop into a symbolic chain often leads to an unmanageably
large graph and does not allow for ``while''-style loops with a variable number
of iterations.

The \emph{Scan} operator is designed to address all of these issues by abstracting the entire
loop into a single node in the graph, a node that communicates with a second
symbolic graph representing computations inside the loop. Without going into copious
detail, we present a list of the advantages of our strategy and refer to
section~\ref{sec:benchmark_scan} where we empirically demonstrate some of these advantages.
Tutorials available from the Theano website offer a detailed description of the required
syntax as well as example code.

\begin{enumerate}
    \item Scan allows for efficient computation of gradients and implicit ``vector-Jacobian''
        products. The specific algorithm used is \emph{backpropagation through time} \cite{Rumelhart86c},
        which optimizes for speed but not memory consumption.
    \item Scan allows for efficient evaluation of the R-operator (see \cite{pearlmutter94}), required
        for computing quantities such as the Gauss-Newton approximation of Hessian-vector products.
    \item The number of iterations performed by Scan can itself be expressed as a symbolic variable (for example,
        the length of some input sequence) or a symbolically specified condition, in which case Scan behaves as a
        ``do while'' statement. If the number of steps is fixed and equal to 1, the Scan
        node is ``unrolled'' into the outer graph for better performance.
    \item Any loop implemented with Scan can be transparently transferred to a GPU (if the computation
        at each iteration can itself be performed on the GPU).
    \item The body of Scan (which involves computing indices of where to pick input slices and
        where to put the output of each iteration) is implemented with Cython to minimize the overhead
        introduced by necessary bookkeeping between each iteration step.
    \item Whenever possible, Scan detects the amount of memory necessary to carry out an operation:
        it examines intermediate results and makes an informed decision as to whether such
        results are needed in subsequent iterations in order to partially optimize memory reuse. 
	This decision is taken at compilation time.
    \item Loops represented as different Scan instances are merged (given that certain necessary
        conditions are respected, e.g., both loops perform the same number of steps). This 
	aids not only in reducing the overhead introduced by each instance of Scan, but
        also helps optimize the computation performed at each iteration of both loops, e.g.
	certain intermediate quantities may be useful to the body of each individual loop, and
	will be computed only once in the merged instance.
    \item Finally, whenever a computation inside the loop body could be performed outside the loop,
        Scan moves said computation in the main graph. For example element-wise operations are moved
        outside, where, given that they are done by a single call to an Elementwise operations,
        one can reduce overhead. Another example is dot products between a vector and a matrix, which
        can be transformed outside of the loop into a single matrix-matrix multiplication. Such
        optimizations can lead to significant speed improvement and in certain cases to the
        elimination of the Scan node completely.

\end{enumerate}

All of these features make it easier for a user to implement a variety
of recurrent neural networks architectures, and to easily change the
equations of the model without having to derive gradients by hand or worry
about manually optimizing the implementation.

\subsection{R-operator for Hessian-Free optimization}
\label{sec:r-op}

Recent results \citep{Martens11}
proposed a specific pipeline for efficiently implementing truncated Newton-like second-order methods
such as Hessian-Free optimization.
The pipeline relies on the ``R-operator'', introduced by \cite{pearlmutter94}, which is a mathematical
operator that given a function $f(\theta), f:\mathbb{R}^M \to \mathbb{R}^N$, the current
parameter configuration $\theta_t \in \mathbb{R}^M$ and a vector $\gamma \in \mathbb{R}^M$,
efficiently computes the ``Jacobian-vector'' product
$\left(\left.\frac{\partial f}{\partial \theta}\right|_{\theta = \theta_t}\right) \gamma$,
where $\left(\left.\frac{\partial f}{\partial \theta}\right|_{\theta = \theta_t}\right)$
is the Jacobian of the function evaluated at $\theta_t$. For the sake of completeness, we would
mention that the ``R-operator'' evaluates the directional derivative of $f(\theta)$, and
is known in the automatic differentiation community as the \emph{forward mode}.

This operation can be seen analogous to the \emph{backward mode} or backpropagation,
which computes the ``vector-Jacobian'' product
$\eta^T \left(\left.\frac{\partial f}{\partial \theta}\right|_{\theta = \theta_t}\right)$,
where $\eta^T \in \mathbb{R}^N$ is some row vector.

Theano offers efficient computation of both operators by employing the chain rule on the
computational graph, where each operational node knows how to compute the product of its
Jacobian and some vector in an efficient way.

Because the output of any such operation is a symbolic
graph, the computations get further optimized at compilation time. This provides
flexibility in writing down the computations that represent a model, without worrying
about details that would lead to faster gradients, or faster ``Jacobian-vector'' products.
For example, let us consider a complicated model, a recurrent network and the task of
computing the Gauss-Newton approximation of the Hessian times some vector (which lies
at the heart of the Hessian-Free algorithm). A naive implementation would imply at least
three passes through the loop, once for evaluating the function, the second one to
backpropagate the gradient (reverse-mode) and the third time to compute the
``Jacobian-vector'' dot product involved in the equation $\frac{\partial f}{\partial \theta}
\left(\frac{\partial f}{\partial \theta}^T\right)\gamma$.
A more careful implementation however reveals that two passes should be sufficient
(see~\cite{Martens11}). By simply calling
\emph{TT.Lop(f, $\theta$, TT.Rop(f, $\theta$, $\gamma$))}, Theano is able to figure
out the relationship between the different loops, resulting in only two passes.

\subsection{Lazy Evaluation, CVM}
\label{sec:cvm}

When a compiled Theano function is called, a runtime engine orchestrates which
operations should be performed on which data, calling the appropriate
functions in the right order. This was previously implemented as a Python loop,
calling either native Python functions or C functions made available
through a Python module interface, in a pre-determined order (i.e., a forward
traversal of the computational graph, from inputs to outputs). The main
drawback of this approach is that it was impossible to implement lazy evaluation
in the computational graph.

For instance, the ``if-then-else'' construct would always compute the
result of both ``then'' and ``else'' branches, as well as the condition,
before updating its output value. A new runtime, dubbed the ``VM''
(for ``Virtual Machine'', because it drives the execution of small code units),
enables lazy evaluation of such operations, meaning that we evaluate only
branches that are actually necessary for correct computation of the output.

A C implementation of the VM was also added (dubbed the ``CVM''). Beyond the
performance advantage inherent in running the loop itself in C, the CVM also
avoids the performance penalty of returning control to the Python interpreter
after each operation: if a C implementation of a given operation is available,
the CVM will execute it directly without the overhead of a Python function call.
The performance gain is particularly significant for graphs that perform many
operations on relatively small operands. In particular, if all operations used
in a compiled Theano function have C implementations, the entirety of the
CVM's execution will be performed at C speed, returning control to the Python
interpreter only after all outputs have been computed. The CVM is now the
default runtime.

\subsection{More operations implemented in C}

To derive fuller benefit from the existence of the CVM, we have added new
C implementations of existing operations (even when Python implementations
were almost as efficient) in order to avoid context
switches. For instance, matrix-vector dot products on CPU had previously
resulted in a call to a SciPy function that wraps the GEMV routine from BLAS.
We have since added a wrapper in C that calls the GEMV routine directly.

\subsection{Better support for sparse matrices}

In addition to dense tensors, Theano supports sparse matrices based
on SciPy's implementations of compressed sparse row (CSR) and compressed sparse column
(CSC) formats. Support for efficient sparse operations, in particular
operations needed to compute derivatives of sparse operations, has been
greatly improved. The online
documentation\footnote{\url{http://deeplearning.net/software/theano/library/sparse/}}
lists currently supported operations.

Theano supports two kinds of gradient computation through sparse
matrices. ``Regular'' differentiation does not suppose that the sparsity
structure of a matrix at a given time is preserved, and thus a sparse variable
may have a dense gradient. ``Structured'' differentiation considers the
sparsity structure of a matrix as permanent, and the gradient with respect to
that matrix will have the same sparsity structure.

\subsection{Parallelism on CPU}

In the past, not much effort had been put into allowing Theano to
leverage multi-core CPU architectures for parallel execution; development
effort was instead focused on GPU implementations and new automatic
optimizations. Multi-core parallelism was therefore only available
to operations that called into a parallel BLAS implementation.

\citet{Torch-2011} showed that using OpenMP to parallelize the C
implementation of CPU operations can bring substantial speed improvements
with relatively little development effort. We recently added support for
OpenMP-enabled operations in Theano, and used this support to parallelize
2-dimensional convolution.  Adding parallel implementations for other
operations will proceed more rapidly with this infrastructure in place.


\subsection{Asynchronous function calls on GPU}
\label{sec:async_gpu}

When executing CUDA kernels on the GPU, the function call that starts
it does not wait for the execution of the kernel to complete. Instead,
it will merely schedule the kernel to be executed at some point in the
future, allowing the main program to perform other tasks, including
scheduling other GPU kernels. When the
result of the GPU computation is needed, the program can wait for the
end of the kernel to execute, and return its result.

Before release 0.6, Theano always waited for the result of the
kernel computation as soon as it was launched, effectively preventing
the execution of other operations on the CPU during this time. This
approach eases profiling and debugging because at any given time, it is
clear which GPU kernel is currently being executed, and error messages
are retrieved as soon as possible; however, such an approach prohibits
the concurrent use of CPU-based computation, passing up an opportunity
for further speed gains. The new default behaviour of Theano is not to
wait on the result of GPU computation until it is strictly needed.
It is also possible to revert to the previous
behaviour, which is useful for profiling execution time of
the different GPU kernels.

\section{Benchmarks}
\label{sec:benchmarks}

\citet{bergstra+al:2010-scipy} showed that Theano was faster than many
other tools available at the time,
including Torch5. The following year, \citet{Torch-2011} showed that
Torch7 was faster than Theano on the same benchmarks\footnote{\url{https://github.com/jaberg/DeepLearningBenchmarks}}.

Here we briefly introduce Torch7 and evaluate performance of their latest
versions on neural network tasks, using the aforementioned benchmarks. Then,
Section~\ref{sec:benchmark_scan} will compare the performance of Theano against
another package, RNNLM, when training recurrent neural networks.

\subsection{Torch7}

\label{sec:feat_env}
Torch7~\citep{Torch-2011} is advertised as a Matlab-like environment for
machine learning. It aims to ease development of numerical algorithms and
to allow for their fast execution, while also being easy to extend.

Table~\ref{theano-torch7-feature} provides a summary comparison
of the features provided by Torch7 (including the ones inherited
from Lua) and Theano (including the ones coming from Python and
NumPy/SciPy). This section exposes the common features and differences
between Torch7 and Theano.

\begin{table}[t]
\noindent\begin{minipage}{\textwidth}

\caption{Features comparison between Lua/Torch7 and Python/Theano. The
first section shows common or comparable features. Second and
third part contains Theano's and Torch7's strengths.}

\newcommand{\tablefootnote}[1]{\footnote{\scriptsize{#1}}}
\label{theano-torch7-feature}
\begin{center}
\small{
\begin{tabularx}{\textwidth}{XXX}
\multicolumn{1}{c}{\bf Features} &\multicolumn{1}{c}{\bf Lua/Torch7} &\multicolumn{1}{c}{\bf Python/Theano} \\
\noalign{\smallskip} \hline \noalign{\smallskip}
Scripting language & \yes & \yes \\
Fast execution speed & \yes & \yes \\
Optimized BLAS, LAPACK & \yes & \yes \\
Plotting Environment & \yes & \yes via matplotlib \\
GPU & \yes float only & \yes float only \\
Easy call to C functions & \yes Natively with Lua & \yes via Cython\tablefootnote{\url{http://www.cython.org/}}, \texttt{ctypes}, etc. \\
OS & Linux, MacOS X, FreeBSD
                             & Linux, MacOS X, Windows\\
Public development & \yes on GitHub\tablefootnote{\url{https://github.com/andresy/torch}}
                             & \yes on GitHub\tablefootnote{\url{https://github.com/Theano/Theano}}\\
Unit tests & \yes & \yes Buildbot\tablefootnote{\url{https://groups.google.com/group/theano-buildbot}}, Travis-CI\tablefootnote{\url{http://travis-ci.org/#!/Theano/Theano}} \\
Used in published research & \yes & \yes \\
Used at companies & NEC & Google, Yahoo!, Card.io, startups \\

\noalign{\smallskip} \hline \noalign{\smallskip}
Sparse matrices & \no & \yes \\
Symbolic differentiation & Non-symbolic NN gradient & \yes \\
Differentiation over loop & \no & \yes Scan \\
R-operator & \no & \yes For most operations \\
Automatic graph optimization & \no & \yes \\

\noalign{\smallskip} \hline \noalign{\smallskip}
Parallel functions & \yes OpenMP widely used & Only in BLAS and Conv2D \\
Embeddable in a C app.\ & \yes & \no \\
Informative error messages & \yes & Not always \\

\noalign{\smallskip} \hline \noalign{\smallskip}
\end{tabularx}
}
\end{center}
\end{minipage}
\end{table}

\subsubsection{Common features shared by Torch7 and Theano}

Theano and Torch7 are two computing frameworks that were developed for
the machine learning community, to make it easier to quickly implement
and test new mathematical models and algorithms, without giving up
the execution speed that a manually-optimized implementation would
provide.  Both are the foundation of machine learning specific packages
or projects, notably for neural networks and unsupervised learning.

Like Theano, Torch7 is based on a scripting language (Lua), uses
heavily-optimized scientific computation libraries (for instance, BLAS
and LAPACK for linear algebra computations), and internal modules
written in C/C++, for the sections where execution speed is critical.
It also has the capability of running parallel computation on multi-core
CPUs (via OpenMP), and on GPUs via CUDA.

Both have access to a Matlab-like environment: Torch7 includes modules
for tensor manipulations and plotting, while Theano benefits from
various external Python libraries to perform those tasks (notably SciPy~\citep{scipy-2001},
NumPy~\citep{numpy-2007}, matplotlib~\citep{hunter-matplotlib-2007},
IPython~\citep{perez-ipython-2007}).
\subsubsection{Main differences}

Some of Torch7's strengths stem from Lua's advantages over Python: lower
interpreter overhead, simpler integration with C code, easy embedding
in a C application. In particular, since the overhead of calling a
function and executing C code are lower, higher performance will result
in the case of simple functions (that do not perform large amounts of computation) and
functions that process only a small quantity of data at a time.

Parallelism on multi-core CPUs is another important feature of Torch7,
as it was designed to use Open Multi-Processing (OpenMP) parallel
directives, notably in the tensor and neural network modules. The
potential for CPU parallelization (outside calls to BLAS) in Theano has
only started to be explored.

Theano's distinguishing feature is its powerful engine for
graph optimization and symbolic differentiation, mentioned in
Section~\ref{sec:symbolic_math}. The downside is that users are faced
with a more complex workflow: first, define an abstract mathematical
graph (without values), then optimize and compile it into a callable
function, and finally execute the function.  This additional complexity also
makes it harder to interpret errors that may be raised during the compilation
or execution phase.

\subsection{Benchmarking on Deep Learning Tasks}

\subsubsection{Experimental Setup}

The experiments reported here were conducted on a machine with an Intel
Core i7 CPU 930 @ 2.80GHz, and a nVidia GTX480 GPU\@. The commit ID of
Theano's version was \texttt{254894fac}, Torch7 was \texttt{41d3b8b93}.

As the multi-layer perceptron (MLP) examples in the benchmarks rely
on function calls to a BLAS library, we made sure the same
BLAS library was used for both Torch7 and Theano, in order to ensure a
fair comparison.
We benchmarked the GEMM routine (matrix-matrix dot product, scaling and
accumulation), with matrix sizes large enough that any overhead becomes
negligible,
for a number of OpenMP threads limited to 1 and 4, confirming that both tools
are linked to the same BLAS library, and that controlling the number of
OpenMP threads works as expected.

\begin{figure}[ht]
\begin{center}

  \subfloat[Logistic regression]{
    \centering
    \includegraphics[width=0.45\textwidth]{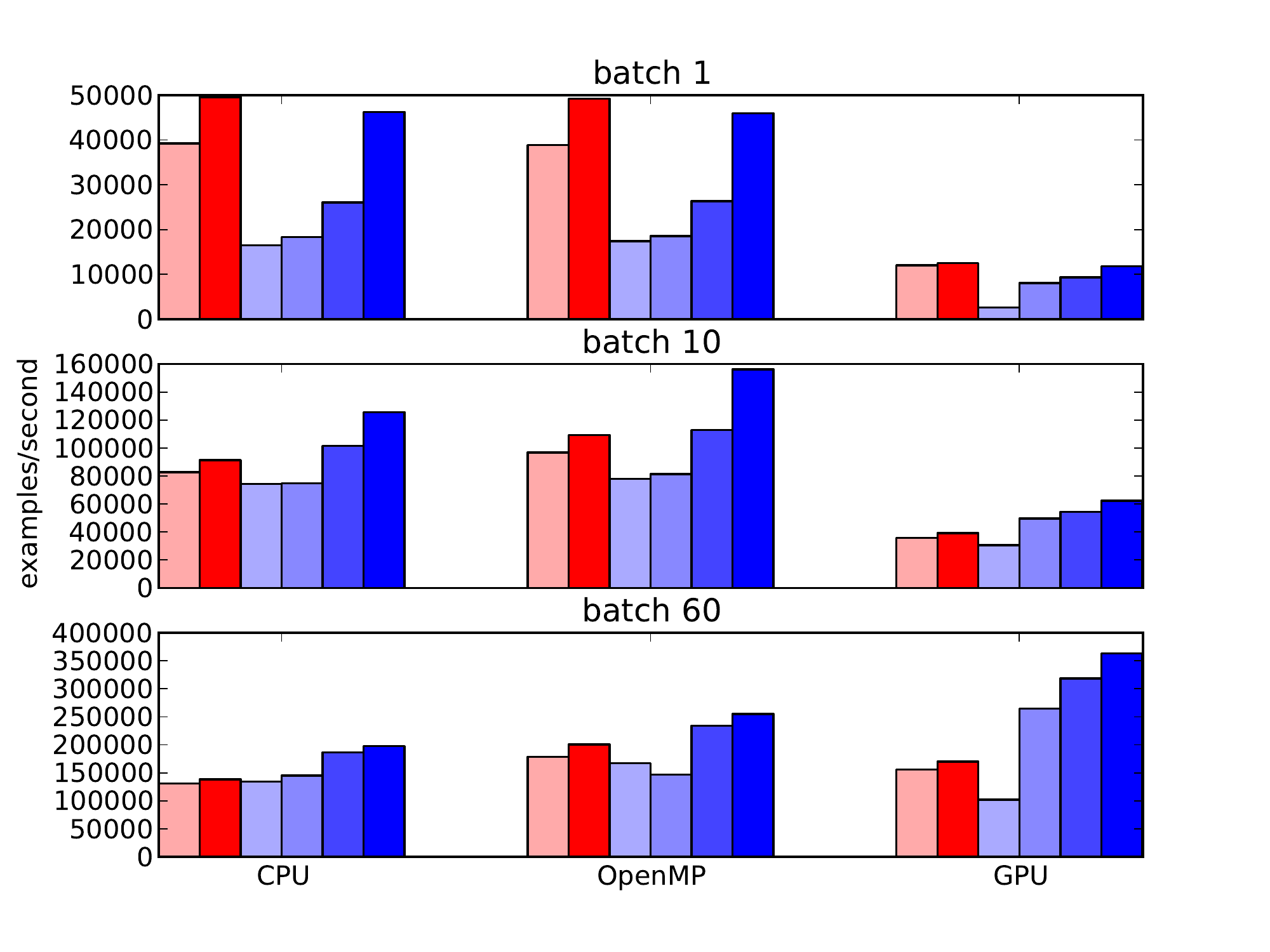}
    \label{fig:mlp0h}
  }~
  \subfloat[Neural network, 1 hidden layer with 500 units]{
    \centering
    \includegraphics[width=0.45\textwidth]{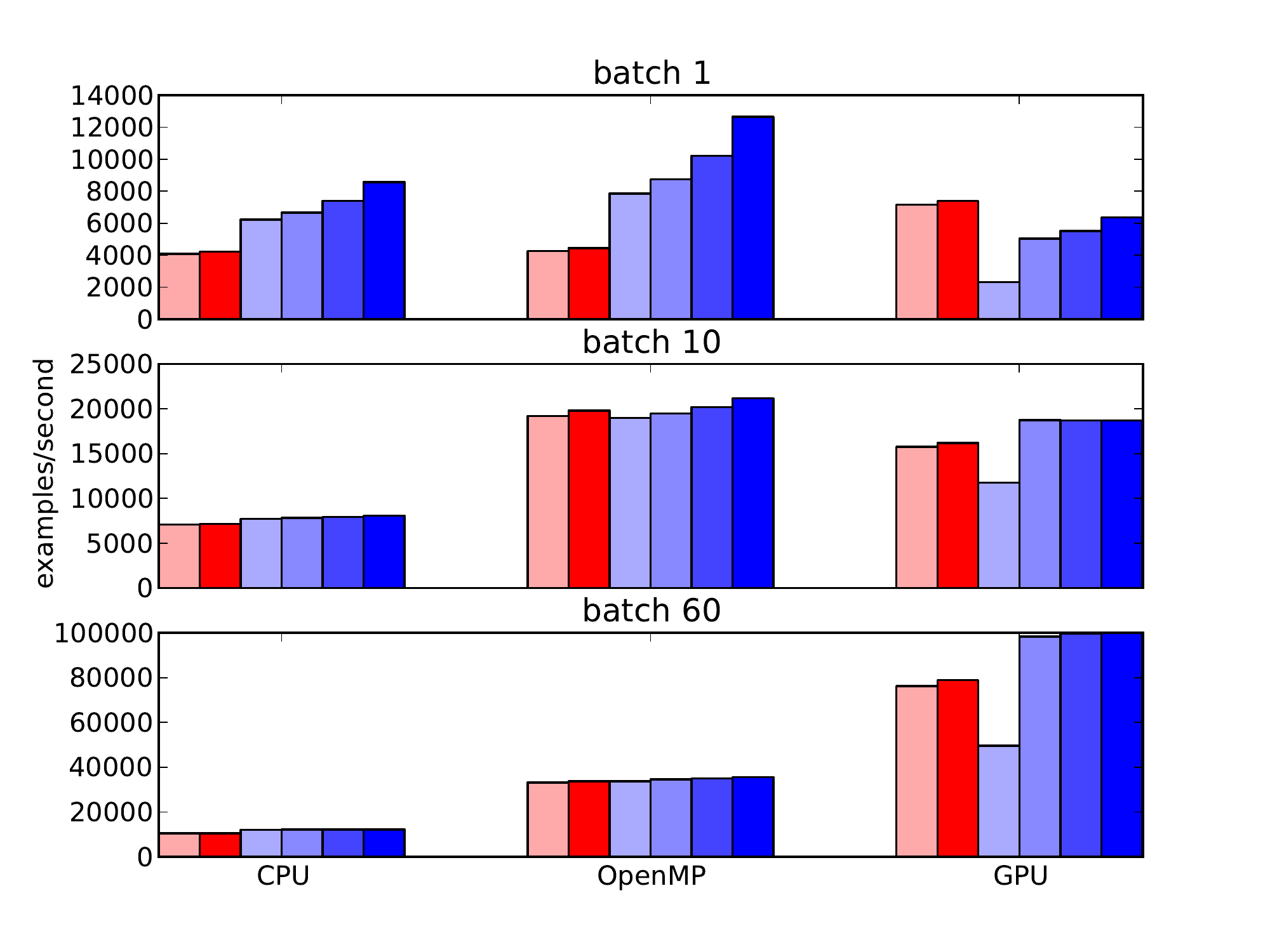}
    \label{fig:mlp1h}
  }\\
  \subfloat[Deep neural network, 3 hidden layers with 1000 units each]{
    \centering
    \includegraphics[width=0.45\textwidth]{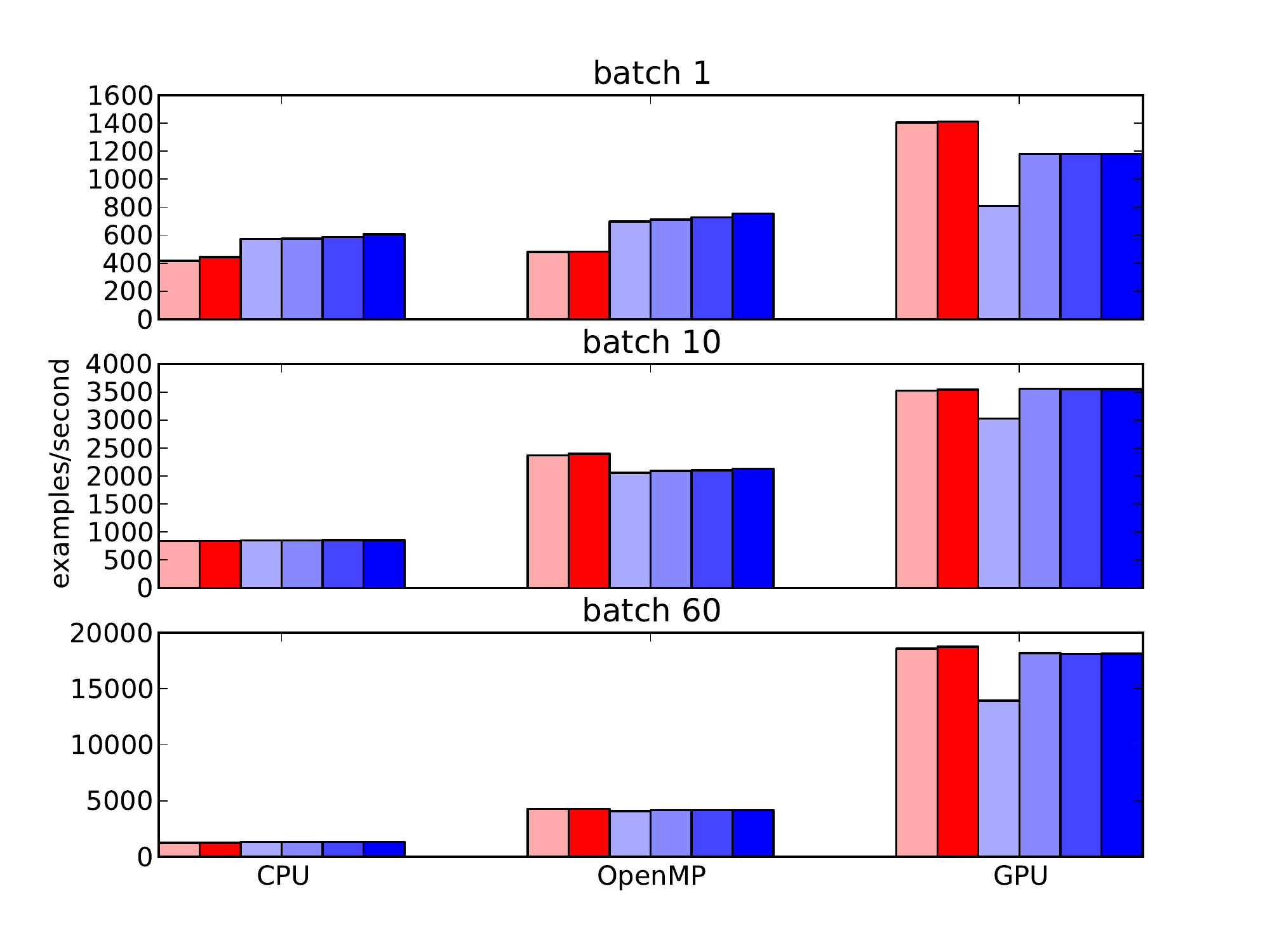}
    \label{fig:mlp3h}
  }~
  \includegraphics[width=0.45\textwidth]{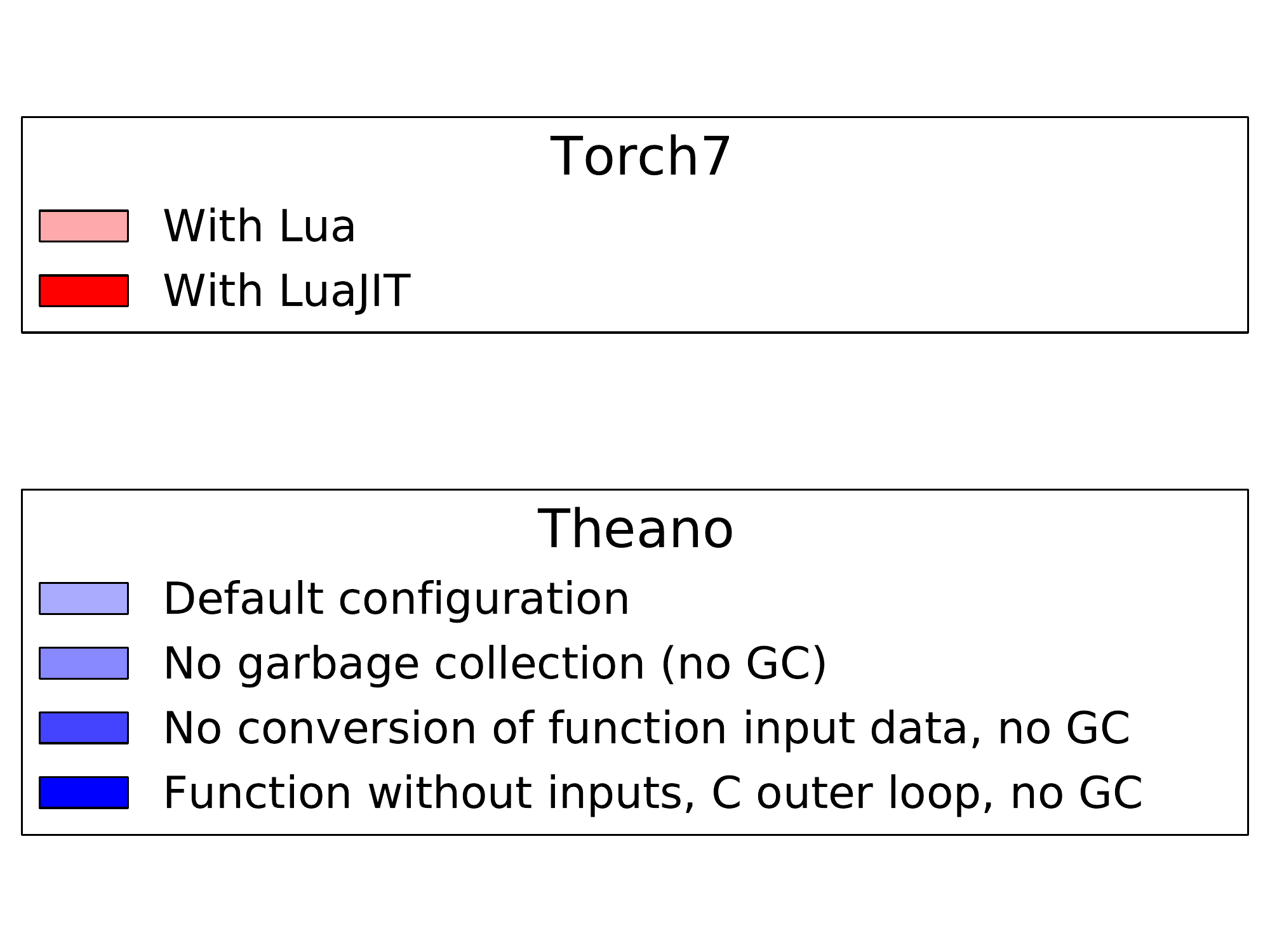}

\end{center}

\caption{Benchmarks of Torch7 (red, left) vs. Theano (blue, right), for training neural networks.
The bars represent examples per second, higher is better.
For each architecture (a), (b), and (c), top figures represent
stochastic gradient descent, middle and bottom ones use mini-batches of
sizes (resp.) 10 and 60.
The group of bars on the left shows performance on CPU with one thread only,
the middle group shows the parallelized version with OpenMP, using 4 CPU threads,
and the right-most show performance on GPU.
}

\label{fig:benchmark}
\end{figure}

\subsubsection{How to boost Theano's performance}
\label{sec:speed}

In Figure~\ref{fig:benchmark}, the left-most blue bar (lightest shade
of blue) in each of the bar groups shows the performance of a Theano
function with the default configuration.  That default configuration
includes the use of the CVM (section~\ref{sec:cvm}), and asynchronous
execution of GPU ops (section~\ref{sec:async_gpu}).  This section
shows ways to further speed up the execution, while trading off other
features.

\paragraph{Disabling garbage collection}

We can save time on memory allocation by disabling garbage
collection of intermediate results. This can be done by using the linker
\verb'cvm_nogc'. In this case, the results of intermediate computation
inside a Theano function will not be deallocated, so during the next
call to the same function, this memory will be reused, and new memory
will not have to be allocated.  This increases memory usage, but speeds
up execution of the Theano function.

In Figure~\ref{fig:benchmark}, the second-to-left blue bar shows the
impact of disabling garbage collection. It is most important on the
GPU, because the garbage-collection mechanism forces a synchronization
of the GPU threads, largely negating the benefits of asynchronous kernel
execution.

\paragraph{Removing overhead of data conversion}

When a Theano function is called and the data type of the provided
input is different from the expected one, a silent conversion is
automatically done (if no precision would be lost).
For instance, a list of integers will be converted into
a vector of floats, but floats will not be converted into integers (an
error will be raised).

This is a useful feature, but checking and converting
the input data each time the function is called can be detrimental
to performance.  It is now possible to disable these checks and
conversions, which gives better performance when the input data is
actually of the correct type. If the input data would actually need to
be converted, then some exceptions due to unexpected data types will
be raised during the execution. To disable these checks, simply set
the \verb+trust_input+ attribute of a compiled Theano function to {\tt
True}. The third blue bar on Figure~\ref{fig:benchmark} shows the speed
up gained with this optimization, including the \verb'cvm_nogc'
optimization.

\paragraph{Executing several iterations at once}

When a Theano function does not have any explicit input (all the
necessary values are stored in shared variables, for instance),
we can save even more overhead by calling its {\tt fn} member:
\verb`f.fn()`. It is also possible to call the same function multiple
consecutive times, by calling \verb`f.fn(n_calls=N)`, saving more
time. This allows to bypass the Python loop, but it will only return
the results of the last iteration. This restriction means that it
cannot be used everywhere, but it is still useful in some cases, for
instance training a learning algorithm by iterating over a data set,
where the important thing is the updates to the parameters, not the
function's output.  The performance of this last way of calling a Theano
function is shown in the right-most, dark blue bar.

\subsubsection{Results}

Figure~\ref{fig:benchmark} shows speed results (in example per
second, higher is better) on three neural network learning tasks,
which consists in 10-class classification of a 784-dimensional
input. Figure~\ref{fig:mlp0h} shows simple logistic regression,
Figure~\ref{fig:mlp1h} shows a neural network with one layer of
500 hidden units, and Figure~\ref{fig:mlp3h} shows a deep neural
network, with 3 layers of 1000 hidden units each. Torch7 was
tested with the standard Lua interpreter (pale red bars), and
LuaJIT\footnote{\url{http://luajit.org/}}, a Lua just-in-time compiler
(darker red bars); Theano was tested with different optimizations
(shades of blue), described in section~\ref{sec:speed}.

When not using mini-batches, on CPU, Theano beats Torch7 on the models
with at least one hidden layer, and even benefits from BLAS parallel
implementation. On the logistic regression benchmark, Torch7 has the
advantage, due to the small amount of computation being done in each
call (executing several iterations at once help, but not enough to beat
LuaJIT). Torch7 also has an edge over Theano on the GPU, when the batch
size is one.

When using mini-batches, whether of size 10 or 60, Theano is faster
than Torch7 on all three architectures, or has an equivalent speed. The
difference vanishes for the most computationally-intensive tasks, as the
language and framework overhead becomes negligible.

\subsection{Benchmarking on Recurrent Neural Networks}
\label{sec:benchmark_scan}

\begin{figure}[ht]
\begin{center}
\includegraphics[width=0.8\textwidth]{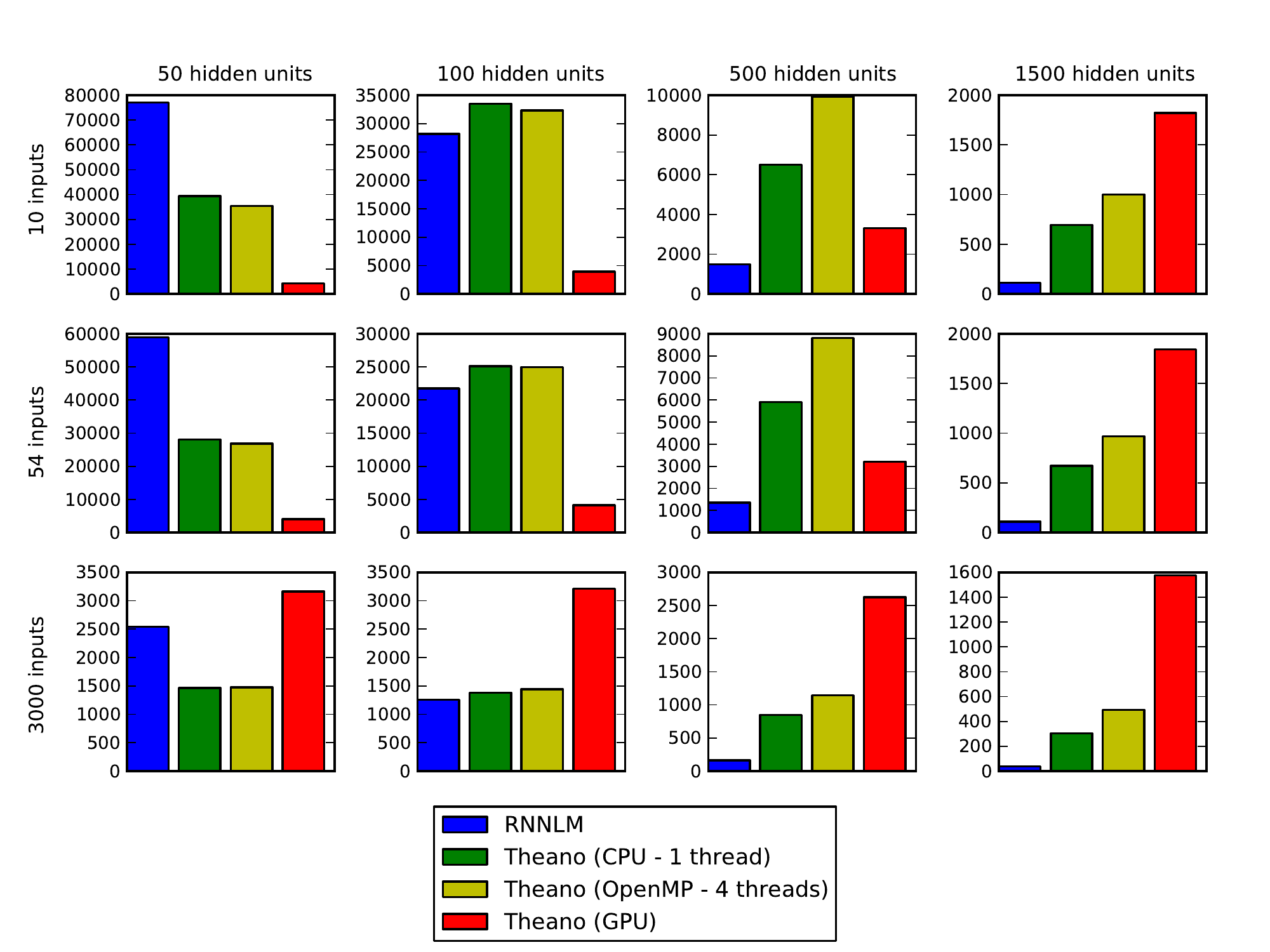}
\end{center}

\caption{Benchmarks of RNNLM vs. Theano on recurrent neural
networks. Reported numbers are sequence elements per second (bigger is
better). The number of input units and output units is the same as the
number of input units.}

\label{fig:RNN}
\end{figure}

In Figure~\ref{fig:RNN}, we present a benchmark
of a simple recurrent network, on Theano and
RNNLM\footnote{\url{http://www.fit.vutbr.cz/~imikolov/rnnlm/}}, a C++
implementation of recurrent networks for language modeling. They
were done with a batch size of 1, which is customary with recurrent
neural networks.

While RNNLM is faster than Theano on smaller
models, Theano quickly catches up for bigger sizes, showing that Theano
is an interesting option for training recurrent neural networks in a
realistic scenario.  This is mostly due to overhead in Theano, which is
a drawback of the flexibility provided for recurrent models.

%
%
%

\section{Conclusion}
\label{sec:conclusion}

We presented recent additions to Theano, and showed how they make it a
more powerful tool for machine learning software development, and allow
it to be faster than competing software in most cases, on different
benchmarks.

These benchmarks aim at exposing relative strengths of existing
software, so that users can choose what suits their needs best. We also
hope such benchmarks will help improving the available tools, which can
only have a positive effect for the research community.

\section*{Acknowledgments}

We would like to thank the community of users and developpers of Theano
for their support, NSERC and Canada Research Chairs for funding, and
Compute Canada and Calcul Qu\'ebec for computing resources.

\small{
\bibliography{strings-shorter,ml,aigaion-shorter,software}

\begin{thebibliography}{}

\bibitem[Bergstra {\em et~al.}(2010)Bergstra, Breuleux, Bastien, Lamblin,
  Pascanu, Desjardins, Turian, Warde-Farley, and
  Bengio]{bergstra+al:2010-scipy}
Bergstra, J., Breuleux, O., Bastien, F., Lamblin, P., Pascanu, R., Desjardins,
  G., Turian, J., Warde-Farley, D., and Bengio, Y. (2010).
\newblock Theano: a {CPU} and {GPU} math expression compiler.
\newblock In {\em Proceedings of the Python for Scientific Computing Conference
  ({SciPy})\/}.
\newblock Oral Presentation.

\bibitem[Bergstra {\em et~al.}(2011)Bergstra, Bastien, Breuleux, Lamblin,
  Pascanu, Delalleau, Desjardins, Warde-Farley, Goodfellow, Bergeron, and
  Bengio]{bergstra+all-Theano-NIPS2011}
Bergstra, J., Bastien, F., Breuleux, O., Lamblin, P., Pascanu, R., Delalleau,
  O., Desjardins, G., Warde-Farley, D., Goodfellow, I., Bergeron, A., and
  Bengio, Y. (2011).
\newblock Theano: Deep learning on gpus with python.
\newblock In {\em Big Learn workshop, NIPS'11\/}.

\bibitem[Collobert {\em et~al.}(2011)Collobert, Kavukcuoglu, and
  Farabet]{Torch-2011}
Collobert, R., Kavukcuoglu, K., and Farabet, C. (2011).
\newblock Torch7: A matlab-like environment for machine learning.
\newblock In {\em Big{L}earn, {NIPS} {W}orkshop\/}.

\bibitem[Hunter(2007)Hunter]{hunter-matplotlib-2007}
Hunter, J.~D. (2007).
\newblock Matplotlib: A 2d graphics environment.
\newblock {\em Computing in Science and Engineering\/}, {\bf 9}(3), 90--95.

\bibitem[Jones {\em et~al.}(2001)Jones, Oliphant, Peterson, {\em
  et~al.}]{scipy-2001}
Jones, E., Oliphant, T., Peterson, P., {\em et~al.} (2001).
\newblock {SciPy}: Open source scientific tools for {Python}.

\bibitem[Martens and Sutskever(2011)Martens and Sutskever]{Martens11}
Martens, J. and Sutskever, I. (2011).
\newblock Learning recurrent neural networks with hessian-free optimization.
\newblock In L.~Getoor and T.~Scheffer, editors, {\em Proceedings of the 28th
  International Conference on Machine Learning (ICML-11)\/}, ICML '11, pages
  1033--1040, New York, NY, USA. ACM.

\bibitem[Mikolov {\em et~al.}(2011)Mikolov, Deoras, Kombrink, Burget, and
  Cernocky]{Mikolov-Interspeech-2011}
Mikolov, T., Deoras, A., Kombrink, S., Burget, L., and Cernocky, J. (2011).
\newblock Empirical evaluation and combination of advanced language modeling
  techniques.
\newblock In {\em Proc. 12th annual conference of the international speech
  communication association (INTERSPEECH 2011)\/}.

\bibitem[Oliphant(2007)Oliphant]{numpy-2007}
Oliphant, T.~E. (2007).
\newblock Python for scientific computing.
\newblock {\em Computing in Science and Engineering\/}, {\bf 9}, 10--20.

\bibitem[Pearlmutter(1994)Pearlmutter]{pearlmutter94}
Pearlmutter, B.~A. (1994).
\newblock Fast exact multiplication by the hessian.
\newblock {\em Neural Computation\/}, {\bf 6}, 147--160.

\bibitem[P\'{e}rez and Granger(2007)P\'{e}rez and Granger]{perez-ipython-2007}
P\'{e}rez, F. and Granger, B.~E. (2007).
\newblock {IPython}: A system for interactive scientific computing.
\newblock {\em Computing in Science and Engineering\/}, {\bf 9}(3), 21--29.

\bibitem[Rumelhart {\em et~al.}(1986)Rumelhart, Hinton, and
  Williams]{Rumelhart86c}
Rumelhart, D.~E., Hinton, G.~E., and Williams, R.~J. (1986).
\newblock Learning internal representations by error propagation.
\newblock volume~1, chapter~8, pages 318--362. MIT Press, Cambridge.

\end{thebibliography}
\bibliographystyle{natbib}
}

\end{document}